\documentclass[twocolumn,superscriptaddress,showpacs,nofootinbib]{revtex4}

\usepackage{graphicx}
\usepackage{dcolumn}
\usepackage{bm}
\usepackage{hyperref}
\usepackage{textcomp}
\usepackage{amsmath}
\usepackage{multirow}

\begin{document}


\title{Optical Scattering Lengths in Large Liquid-Scintillator Neutrino Detectors}

\author{M. Wurm}\email[Corresponding author, e-mail:~]{mwurm@ph.tum.de}
\author{F. von Feilitzsch}
\author{M. G\"oger-Neff}
\author{M. Hofmann}
\affiliation{Physik-Department E15, Technische Universit\"at M\"unchen, James-Franck-Str., D-85748 Garching, Germany}
\author{T. Lachenmaier}
\affiliation{Excellence Cluster Universe, Technische Universit\"at M\"unchen, Boltzmannstr. 2, D-85748 Garching, Germany}
\author{T. Lewke}
\affiliation{Physik-Department E15, Technische Universit\"at M\"unchen, James-Franck-Str., D-85748 Garching, Germany}
\author{T. \surname{Marrod\'an Undagoitia}}
\affiliation{Physik-Department E15, Technische Universit\"at M\"unchen, James-Franck-Str., D-85748 Garching, Germany}
\affiliation{Physik-Institut, Universit\"at Z\"urich, Winterthurstr.\,189, CH-8057 Z\"urich, Switzerland}
\author{Q. Meindl}
\author{R. M\"ollenberg}
\author{L. Oberauer}
\author{W. Potzel}
\author{M. Tippmann}
\author{S. Todor}
\affiliation{Physik-Department E15, Technische Universit\"at M\"unchen, James-Franck-Str., D-85748 Garching, Germany}
\author{C. Traunsteiner}
\affiliation{Excellence Cluster Universe, Technische Universit\"at M\"unchen, Boltzmannstr. 2, D-85748 Garching, Germany}
\author{J. Winter}
\affiliation{Physik-Department E15, Technische Universit\"at M\"unchen, James-Franck-Str., D-85748 Garching, Germany}

\date{\today}

\begin{abstract}

For liquid-scintillator neutrino detectors of kiloton scale, the transparency of the organic solvent is of central importance. The present paper reports on laboratory measurements of the optical scattering lengths of the organic solvents PXE, LAB, and Dodecane which are under discussion for next-generation experiments like SNO+, \textsc{Hanohano}, or LENA. Results comprise the wavelength range from 415 to 440\,nm. The contributions from Rayleigh and Mie scattering as well as from absorption/re-emission processes are discussed. Based on the present results, LAB seems to be the preferred solvent for a large-volume detector.

\end{abstract}



\maketitle

\section{Introduction}
\label{SecIntrod}

In recent years, large-volume liquid-scintillator detectors have made important contributions to low-energy neutrino physics \cite{kam02,cho03,bx07be7}. At present, typical target masses range from several hundred tons to kilotons, corresponding to spherical volumes of more than 5 meters in radius \cite{sue04kl, bx08det}. Scintillation light generated in the bulk of the detector has to cross several meters of liquid before arriving at the photomultipliers. As a consequence, the transparency of the organic solvent plays an important role for the signal interpretation in present-day neutrino experiments.

The importance of optimizing the transmission of scintillation light in organic solvents will dramatically increase in the next generation of liquid-scintillator detectors: the up-coming SNO+ experiment (1\,kt target mass) \cite{che05} and especially the planned \textsc{Hanohano} (10\,kt) \cite{lea08} and LENA (50\,kt) \cite{mar08} neutrino observatories demand efficient light transport over 10 to 20 meters. Even given a high molecular purity of the used solvents, this requirement is already near to the optimum transparency that can be obtained under realistic conditions. The natural limit that Rayleigh scattering off the solvent molecules imposes on the attenuation length is typically of the order of 30\,m. 

The present work describes measurements of the optical scattering lengths in a number of organic solvents that are currently discussed as candidates for upcoming experiments. The results complement preceding measurements of the attenuation length that reflects the general transparency \cite{bx00sci} as it includes both the effects of light scattering and light absorption. However, both parameters are mandatory for a correct description of three-dimensional light transport.

The analysis of the data makes it necessary to distinguish between a variety of microscopic processes that can contribute to the scattering and absorption in an organic liquid. Sect.\,\ref{SecLigPro} gives a comprehensive overview of the most important mechanisms and their effects on macroscopic light propagation. The laboratory setup used for the actual measurements is described in Sect.\,\ref{SecExpSet}, data analysis is treated in Sect.\,\ref{SecDatAna}. The results as well as their interpretation are presented in Sects.\,\ref{SecExpRes} and \ref{SecConclu}.

\section{Light Propagation}
\label{SecLigPro}

Light traveling through an organic liquid scintillator can either be deflected from its path or be absorbed by molecules impeding its way. The underlying microscopic processes can be distinguished by the direction of the outgoing photon (if there is a final-state photon) and its polarization. In general, the individual contributions of different processes to the overall macroscopic light attenuation depend on the molecular properties of the solvent, its purity, and the number density of interaction centers.


The three microscopic processes having influence on the light propagation are the following:
\begin{enumerate}
\item Rayleigh scattering off the bound electrons of molecules in the solvent \cite{bohren}. 
\item Mie scattering from dirt or dust particles suspended in the liquid \cite{born,hulst}.
\item Absorption of the light by molecules, which is either re-emitted or converted into excitation modes invisible to the photomultipliers (like infrared light or heat) \cite{bx00sci}.
\end{enumerate}

Only if the light is fully absorbed, it is completely lost for detection, while in all other scenarios merely the propagation direction changes. Therefore, absorption-reemission processes can be counted as a form of scattering for all practicable purposes. Raman scattering describing inelastic scattering by molecules can be neglected as it is usually several orders of magnitude weaker than Rayleigh scattering \cite{harris}.


Experimentally, the contributions of these processes can be distinguished by the angular dependence of the scattered light: The differential cross-section of Rayleigh scattering \cite{bohren}

\begin{equation}
\label{EqRaySca1}
\left(\frac{d\sigma}{d\Omega}\right)_\mathrm{ray} \propto \left( \frac{1+\cos^2\theta}{2} \right),
\end{equation}

reflects the full suppression of the polarization component parallel to the direction of incident light for a scattering angle $\theta$=$90^\circ$. The intensity emitted over the whole solid angle is therefore anisotropic, reaching twice the value of that corresponding to $\theta$=90$^\circ$ for scattering directions parallel (or antiparallel) to the initial propagation direction.

Re-emission processes on the other hand are assumed to be fully isotropic as the light is first absorbed by the molecule and the delayed de-excitation is unrelated to the incident light direction.

Depending on the size of the scattering centers, Mie scattering features either Rayleigh-like or highly irregular differential cross-sections, usually with an increased forward-scattering amplitude \cite{born,hulst}. Depending on the material and size of the particles, polarization of the scattered light is more or less pronounced. Possible indications for Mie scattering are an additional contribution of anisotropic scattering that cannot be accounted for by Rayleigh scattering or a pronounced forward-scattering amplitude.

In the present experiment, both the intensity and polarization of the scattered light are measured for several scattering angles. Using these results, it is possible to identify the contributions of Rayleigh scattering, Mie scattering and absorption/re-emission processes by their angular dependences. The total number of scattered photons per unit propagation length $\mathrm d N/\mathrm d x$ can therefore be derived without an actual $4\pi$-detection of the scattered intensity. 


Assuming one-dimensional propagation, $\mathrm d N/\mathrm d x$ can directly be related to the attenuation length $L$ which describes the length after which the number of photons $N(x)$ has dropped to $1/e$ of its initial value $N_{0}$:

\begin{equation}
\label{EqScaLen}
\frac{\mathrm d N}{\mathrm d x} = - Q N(x) \Rightarrow N(x) = N_{0} \mathrm e^{-Q x} = N_{0} e^{-x/L}.
\end{equation}

The factor $Q=n\sigma_\mathrm{tot}$ describes the ratio of scattered light per unit length to the incident number of photons $N(x)$ at point $x$: Here, $n$ is the number density of scattering centers in the direction of light propagation, while $\sigma_\mathrm{tot}$ is the total interaction cross section. It is also possible to derive the partial light propagation lengths $\ell_{i}$  which are related to the scattering ratios $Q_{i}$ of individual processes via $\ell_{i}=1/Q_{i}$. As $Q$=$\sum_iQ_i$=$\sum_i (n \sigma_\mathrm{tot})_i$, the attenuation length $L$ can be written as
\begin{equation}
\label{EqScaSum}
\frac{1}{L} = \frac{1}{\ell_\mathrm{A}}+\frac{1}{\ell_\mathrm{S}};\qquad \frac{1}{\ell_\mathrm{S}}=\frac{1}{\ell_\mathrm{are}}+\frac{1}{\ell_\mathrm{ray}}+\frac{1}{\ell_\mathrm{mie}}.
\end{equation}
The relevant propagation lengths are:\\~\\
\begin{tabular}{ll}
$L$					& Attenuation Length \\
$\ell_\mathrm{A}$	& Absorption Length \\
$\ell_\mathrm{S}$	& Scattering Length \\
$\ell_\mathrm{are}$	& Absorption/Re-Emission Length \\ 
$\ell_\mathrm{ray}$	& Rayleigh Scattering Length \\ 
$\ell_\mathrm{mie}$	& Mie Scattering Length \\ 
$\ell_\mathrm{an}$	& Anisotropic Scattering Length \\ 
$\ell_\mathrm{is}$	& Isotropic Scattering Length \\ 
\end{tabular}\\~\\
For practical reasons, $\ell_\mathrm{are}$ is here included in the overall scattering length $\ell_\mathrm{S}$. In the present analysis, the scattering length $\ell_\mathrm{S}$ is described as a combination of anisotropic ($\ell_\mathrm{an}$) and isotropic ($\ell_\mathrm{is}$) scattering lengths: 
\begin{equation}
\frac{1}{\ell_\mathrm{S}}=\frac{1}{\ell_\mathrm{an}}+\frac{1}{\ell_\mathrm{is}}.
\end{equation}
This will be discussed in detail in Sect.\,\ref{SecDatAna}.






\section{Experimental Setup}
\label{SecExpSet}

\begin{figure}
\centering
\includegraphics[width=0.48\textwidth]{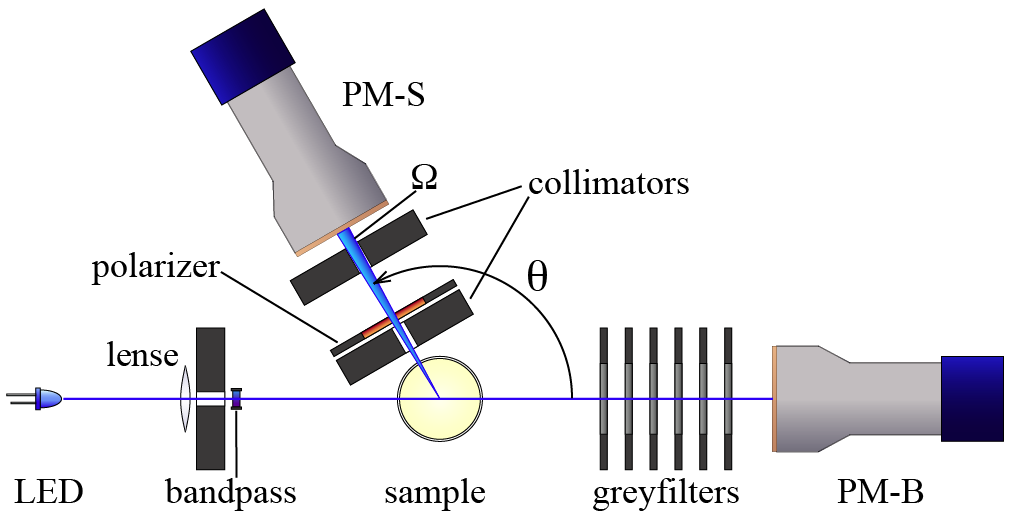}
\caption{Sketch of the experimental setup: The light generated by a LED crosses the sample and is scattered into the photomultiplier PM-S. Collimators define the field of view, while a polarizer is used to disentangle different contributions to the scattering amplitude. The photomultiplier PM-B monitors the beam intensity.}
\label{FigExpSet}
\end{figure}

Fig.\,\ref{FigExpSet} shows a schematic drawing of the laboratory setup. The experiment is based on a collimated light beam of well-defined wavelength that is sent through a sample of liquid scintillator contained in a glass vessel. The number of photons in the beam $N_\mathrm{B}$ is monitored by a photomultiplier PM-B. A second photomultiplier PM-S can be set off-beam at different scattering angles $\theta$ and registers the photon number $N_\mathrm{S}$ scattered by the scintillator. A set of collimators in front of PM-S defines the solid angle $\Omega$ the PM is sensitive to. An additional polarization filter allows to determine to which extent the scattered light is polarized, selecting the linear polarization component $p$: either perpendicular $\perp$ or parallel $\parallel$ to the beam. The result of each measurement is the ratio $q(\theta,p)=N_\mathrm{S}/N_\mathrm{B}$ of the measured photon numbers from which the various propagation lengths will be derived (Sect.\,\ref{SecDatAna}). The whole setup is fully contained in a light-tight box in order to protect the phototubes from external light.

\subsection{Optical Components}

The light source used is an LED emitting between 400 and 460\,nm with the maximum at 430\,nm \cite{led}. A focussing lens together with an aperture inhibits a significant widening of the beam to the photomultiplier PM-B \cite{etl} that is monitoring the beam intensity. A narrow-bandpass filter \cite{nbf} limits the beam spectrum to $\sim$10\,nm around the central wavelength of the filter: Three bandpass filters of $(415.0^{+3.4}_{-3.7})$\,nm, $(430.3^{+4.5}_{-4.2})$\,nm, and $(441.5^{+4.8}_{-3.6})$\,nm are used \cite{wur09phd}.

The scintillator sample is contained in a cylindrical glass beaker of 6\,cm diameter and $\sim$15\,cm in height. This geometry was chosen to minimize light reflection and refraction on the sample surface. In the analysis, the remaining optical effects are taken into account by a Monte Carlo simulation (Sect.\,\ref{SecDatAna}).

In front of the beam-monitoring photomultiplier PM-B, an array of six optical greyfilters \cite{ndf} reduces the light intensity to a level compatible with the high sensitivity of the light detector. The cumulative transmissions through greyfilters and polarizer at three defined wavelengths are reported in Tab.\,\ref{TabCorPar}. The effective aperture of PM-B is 5$\times$5\,cm, i.\,e.\,much larger than the beam diameter, thus avoiding beam shadowing effects.

The scintillator sample as well as the second photomultiplier PM-S are mounted on a rotatable platform. The rotation axis coincides with the axis of the glass beaker. Four scattering angles were used: 75$^\circ$, 90$^\circ$, 105$^\circ$, and 120$^\circ$.

PM-S is of the same type as PM-B. Both PMs are shielded from external magnetic fields by $\mu$-metal foils. The accessible solid angle is well defined by a pair of rectangular collimators. PM-S's field of view is narrowed to the very center of the sample, corresponding to a maximum angular acceptance of $\theta_\mathrm{max}$$\approx$$\pm4^\circ$ and $\varphi_\mathrm{max}$$\approx$$\pm10^\circ$. The angle $\varphi$ describes rotations around the beam axis. In addition, the collimators shield the phototube from the diffuse scattering of light occurring at the entry and exit points of the beam in the beaker glass. To determine the polarization of the scattered light, a linear polarizer \cite{pol} is mounted between the two collimators in front of PM-S.

\subsection{Electronics and DAQ}

A pulse generator \cite{pgr} is used to supply both voltage pulses to the LED and a NIM trigger signal to the data acquisition (DAQ) system described below. The LED is operated with square pulses of $\sim$25\,\textmu s in length and $\sim$2\,V in amplitude.

Both photomultipliers are operated in single-photon mode at a voltage of 1.4\,kV \cite{hvs}. Voltage dividers are mounted on the PM bases. 
 
The DAQ system itself is a combined unit of ADCs and a PC for controlling the DAQ \cite{acq}. A time resolution of 500\,ps (2\,GS/s) and a total voltage range of 50\,mV at a resolution of 10\,bit was applied. Between 0.1 and 20 photons per time gate are registered by the phototubes, equally distributed over the entire gate duration of 25\,\textmu s. This allows an analysis based on the counting of single-photon pulses. In addition, the measured number of photons per gate duration are statistically corrected for pile-up, which is at maximum a 10\,
\% effect. The number of dark counts of the PMs inside the time gate is negligible. 

For each time gate, the number of registered photons is written to an individual histogram for both PM channels. Mean values and standard deviations of the histograms are calculated. The ratio $q=N_\mathrm{S}/N_\mathrm{B}$ of the obtained mean photon numbers serves as input parameter for the further analysis.

The DAQ software is based on LabView \cite{ni}.

\section{Data Analysis}
\label{SecDatAna}

The determination of the scattering length requires a sequence of steps that is executed by a C++/ROOT-based software routine \cite{root}.

For all samples, data is evaluated separately for each wavelength. The experimentally derived intensity ratios $q(\theta,p)$=$N_\mathrm{S}/N_\mathrm{B}$ for four scattering angles $\theta$ and two polarizations $p$ (perpendicular $\perp$ and parallel $\parallel$ to the beam) are used as basic values. Several corrections have to be applied to $q$: 
\begin{itemize}
\item Uncorrelated systematic uncertainties have to be added to the statistical uncertainties (1\%) of the individual values of $q(\theta,p)$. Most prominent are terms considering surface evenness and cleanness of the sample glass (4\%) and variations in the photoefficiency of PM-S due to changes of magnetic stray fields for different values of $\theta$ (7\%).
\item The results of a background measurement $q_\mathrm{bg}=N_\mathrm{S,bg}/N_\mathrm{B,bg}$ using deionized water have to be subtracted from each data point $q$. $q_\mathrm{bg}$ in turn has to be corrected for the residual Rayleigh scattering $\ell_\mathrm{ray}$$\approx$90\,m in the water sample. $\ell_\mathrm{ray}$ is derived from measurements in Super-Kamiokande \cite{SK99phd}, and reduces $q_\mathrm{bg}$ by about 1/3.
\item The relative photoefficiency $\varepsilon$ of PM-S to PM-B is taken into account. $\varepsilon$ has been determined for the used range of wavelengths in an auxiliary measurement (Tab.\,\ref{TabCorPar}). Also the angular dependence of PM-S's efficiency due to magnetic stray fields has been tested. 

\item The transmission coefficients $t_\mathrm{gf}(i)$ of the greyfilters and of the polarization filter ($t_\mathrm{pol}$) are included. The values have been determined by two auxiliary measurements, the first using a slight modification of the described setup, the second employing a portable spectrometer (see Sect.\,\ref{SecExpSet}). Both return compatible results. As the greyfilters reduce the light registered by PM-B, while the polarizer absorbs some of the light scattered in the solid angle of PM-S, the correction factor $T$ (Tab.\,\ref{TabCorPar}) is the quotient of the relative transmission factors:
\begin{eqnarray}
T=\frac{\prod_i t_\mathrm{gf}(i)}{t_\mathrm{pol}}.
\end{eqnarray}
\item Corrections for reflection of the light beam at the glass surface of the sample holder are applied to obtain the ratio of scattered photons $N_\mathrm{S}$ relative to the incident number of photons $N_0$. With $\varrho$ being the relative light loss by surface reflections (Tab.\,\ref{TabCorPar}), the number of photons ($N_\mathrm{B}$) detected by PM-B after crossing the sample is given by $N_\mathrm{B}$=(1-$\varrho)N_0$. The light attenuation by the sample can be neglected.

\end{itemize} 

Applying all corrections, one obtains
\begin{eqnarray}\label{EqCorRat}
  q_\mathrm{cor} = T\varepsilon\,(1-\varrho)\,\left(\frac{N_\mathrm{S}}{N_\mathrm{B}}-\frac{N_\mathrm{S,bg}}{N_\mathrm{B,bg}}\right) = T\varepsilon\,(1-\varrho)\,(q-q_\mathrm{bg}).
\end{eqnarray}

\begin{table}
\begin{center}
\begin{tabular}{lcc}
\hline
Parameter & $\lambda$\,[nm]	&  Value \\ 
\hline
Filter Transmission $T(\lambda)$	& 415 & 5.5$\pm$0.2 \\
 $\qquad[10^{-3}]$ & 430 & 7.8$\pm$0.3 \\
	& 442 & 6.7$\pm$0.3 \\
\hline
Relative photoefficiency & $\varepsilon$		& 0.69$\pm$0.05 \\
Optical reflections & $\varrho$				& 0.083$\pm$0.003 \\
\hline
\end{tabular}
\caption{Corrections applied to the measured ratio of scattered to incident number of photons: The correction $T$ regarding greyfilter and polarizer transmission was measured for three relevant wavelength regions. $\varepsilon$ reflects the relative photoefficiencies of PM-S and PM-B, while $\varrho$ corrects for light losses due to reflections at the sample surfaces (Sec.\,\ref{SecDatAna}).}
\label{TabCorPar}
\end{center}
\end{table}  

In Eq.\,(\ref{EqCorRat}), $N_\mathrm{S}$ is the number of photons measured by PM-S. It is considerably lower than the total number of photons scattered inside the sample: First of all, the field of view is narrowed by the apertures in front of PM-S, which on the one hand reduces the number of scattered photons in the solid angle $\Omega$ allowed by the apertures (Fig.\,\ref{FigExpSet}) and on the other hand introduces an angular dependence on the length of the beam portion seen by PM-S. And lastly, refraction and reflection of the scattered light on the sample holder surface will modify the effective solid angle seen by PM-S \cite{wur09phd}.

Therefore, the most feasible way to correlate $q_\mathrm{cor}$ to a scattering length $\ell$ is the comparison to a Monte Carlo simulation of the experimental geometry. The simulation toolkit \textsc{Geant4} \cite{g4} has been used to retrieve effective values $q_\mathrm{MC}(\theta)$ that represent the ratio of $N_\mathrm{S}$ to $N_\mathrm{0}$ assuming isotropic scattering. The input scattering length $\ell_\mathrm{MC}$ is 1\,m. Therefore, the ratio of the experimental value of the scattering length(s) to the MC length corresponds to the ratio of the measured value $q_\mathrm{cor}$ and the simulated value $q_\mathrm{MC}$. The ratio $q_\mathrm{MC}(\theta)$ takes solid angle, PM-S position and the influence of the glass and sample surfaces into account. The ratio $Q(\theta)=q_\mathrm{cor}(\theta)/(q_\mathrm{MC}(\theta)\ell_\mathrm{MC})$ is independent of the solid angle and has already the dimension of an inverse scattering length.

In total, eight values are retrieved for $Q(\theta,p)$ corresponding to the four angles ($\theta = 75^\circ$, 90$^\circ$, 105$^\circ$, and 120$^\circ$) and the two polarization states $p$ used in the experiment. Scattering in the sample is approximated by two processes: a polarized anisotropic contribution $\mathcal A_{p}(\theta)$ with an intensity distribution $f_\mathrm{an}(\theta)=\frac{1}{2}(1+\cos^2\theta)$ of Eq.\,(\ref{EqRaySca1}); and an unpolarized isotropic component $\mathcal I$ that represents mainly absorption-reemission processes with $f_\mathrm{is}(\theta)=1$. Contributions from Mie scattering following an individual angular distribution can be neglected (Sect.\,\ref{SecExpRes}). Due to the different dependences on $\theta$ and $p$, the contributions of the two processes in the sample can be determined by fitting the sum $\mathcal A_{p}(\theta)+\mathcal I$ to the measured values $Q(\theta,p)$:
\begin{eqnarray}
\label{EqScaFit}
Q_\perp(\theta) & \stackrel{!}{=} & \mathcal A_\perp(\theta)+\mathcal I  =  \frac{1}{2}Q_\mathrm{an}+\frac{1}{2}Q_\mathrm{is} \\
Q_\parallel(\theta) & \stackrel{!}{=} & \mathcal A_\parallel(\theta)+\mathcal I  =  \frac{\cos^2\theta}{2}Q_\mathrm{an}+\frac{1}{2}Q_\mathrm{is} \nonumber
\end{eqnarray}
$Q_\mathrm{an}$ and $Q_\mathrm{is}$ are fit parameters and represent the contributions of anisotropic and isotropic scattering summed for both polarization states $p$.

\begin{figure}
\centering
\includegraphics[width=0.48\textwidth]{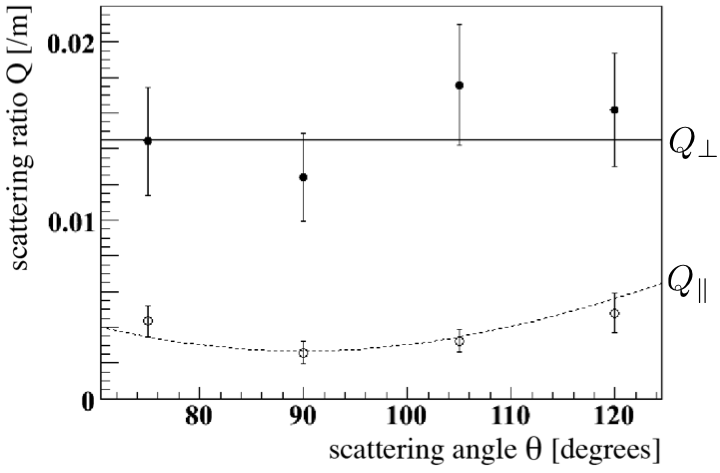}
\caption{Data points $Q(\theta)$ in dependence of the scattering angle $\theta$. Filled (open) circles represent perpendicular (parallel) polarization. The two lines correspond to fits to the data points, parameterized according to Eqs.\,(\ref{EqScaFit}). Data corresponds to a Dodecane sample at 415\,nm wavelength.}
\label{FigFitDat}
\end{figure}

Fig.\,\ref{FigFitDat} shows the data points $Q(\theta,p)$ and the fitted curves for the case of Dodecane (C12) and a wavelength of 415\,nm. Filled circles correspond to perpendicular, open ones to parallel polarization. The isotropic scattering length $\ell_\mathrm{is}$ is the inverse of $Q_\mathrm{is}$ (Sect.\,\ref{SecLigPro}).

While possible in most cases, for some samples the scattering of the data points is too large to retrieve a fit with a reasonable $\chi^2$-value. Under these circumstances, the fit is replaced by an alternative analysis using mean values: The difference in the mean perpendicular ratio  $\langle Q_\perp \rangle$ and the mean parallel ratio $\langle Q_\parallel \rangle$ is due to anisotropic scattering only (Eq.\,(\ref{EqScaFit})). Averaging over all scattering angles $\theta$, the ratios of isotropic and anisotropic scattering can be written as:

\begin{equation} \label{EqMeaRes}
 Q_\mathrm{an} = \frac { 2 \langle Q_\perp\rangle - 2\langle Q_\parallel \rangle } {1-\langle\cos^2\theta\rangle };~
 Q_\mathrm{is} =  \frac { 2 \langle Q_\parallel\rangle - 2\langle Q_\perp \rangle \langle \cos^2\theta\rangle } {1-\langle\cos^2\theta\rangle }
\end{equation}

The uncertainties of the results obtained in this way are usually larger but in good agreement with the results of the fit to the data points.

As the effective ratios $q_\mathrm{MC}$ have been calculated assuming an isotropic scattering, $\ell_\mathrm{is}$ corresponds to the inverse value of $Q_\mathrm{is}$. However, an additional factor of 4/3 appears in the case of anisotropic scattering, as the integral over the angular dependence function $f_\mathrm{an}(\theta)$ that has been implicitly used in Eq.\,\ref{EqScaFit} is only 0.75\,$\pi$ (while the integral of $f_\mathrm{is}(\theta)$ is $\pi$). Therefore,
\begin{eqnarray}
\ell_\mathrm{is} &=& \frac{1}{Q_\mathrm{is}},\nonumber\\
\ell_\mathrm{an} &=& \frac{\int{f_\mathrm{is}(\theta)\mathrm d \theta}}
                      {\int{f_\mathrm{an}(\theta)\mathrm d \theta}}\frac{1}{Q_\mathrm{an}} =
\frac{4}{3}\frac{1}{Q_\mathrm{an}},\nonumber\\
\frac{1}{\ell_\mathrm{S}} &=& \frac{1}{\ell_\mathrm{is}} +  \frac{1}{\ell_\mathrm{an}}. \nonumber
\end{eqnarray}

\section{Experimental Results}
\label{SecExpRes}

The results reported here are from two series of measurements: The first one covers a large variety of solvent samples, while the incident light wavelength is fixed at 430\,nm. The second series investigated the wavelength dependence of a subgroup of solvents by taking measurements at three different wavelengths. 

\subsection{Scattering Lengths at 430\,nm}
\label{SubsRes430}

The measurements were performed at a wavelength of 430\,nm at which scintillation light is well transmitted through the solvent. A broad selection of materials has been screened, including phenylxylylethane 'PXE', pseudocumene 'PC', dodecane 'C12' from two different manufacturers (\textsc{'sa'} for Sigma Aldrich, \textsc{'ac'} for Alway Chem), and three brands of linear alkylbenzene 'LAB' produced by Petresa (\textsc{'p500'}, \textsc{'p550'}, and \textsc{'550q'}). In addition, the solvent cyclohexane 'CX' has been tested, as it is a standard for spectroscopic measurements. The sample of PXE was characterized before and after purification by column chromatography using aluminum oxide. CAS-numbers and material purities of the samples are listed in Tab.\,\ref{TabSciMan} together with the manufacturers and product numbers. 

\begin{table}
\begin{center}
\begin{tabular}{llcl}
\hline
 \textbf{Abbr.} & \textbf{CAS$\#$} & \textbf{Purity} & \textbf{Manufacturer, Product}\\
\hline
PXE				& 6196-95-8	& $>$0.97 	& Dixie, POXE\\
\hline
LAB	\textsc{p500}	& 67774-74-7	& 0.992		& Petresa, Petrelab 500\\
LAB	\textsc{p550}	&			&			& Petresa, Petrelab 550\\
LAB	\textsc{550q}	&			&			& Petresa, Petrelab 550 Q\\
\hline
C12 \textsc{sa}		& 112-40-3	& $\geq$0.98	& Fluka, SA44020 \\
C12 \textsc{ac}		&			&			& Alway Chem \\
\hline
PC				& 95-63-6		& $>$0.98		& Merck, SA814505 \\
\hline
CX				& 110-82-7	& 0.9999		& Merck, Uvasol, SA102822\\
\hline
\end{tabular}
\caption{Overview of the CAS numbers, molecular purities, product names and manufacturers of the used solvents and solutes.}
\label{TabSciMan}
\end{center}
\end{table} 

Tab.\,\ref{TabSerOne} summarizes the results of the analysis. Both the anisotropic and isotropic scattering length $\ell_\mathrm{an}$ and $\ell_\mathrm{is}$ are shown. The total scattering length $\ell_\mathrm{S}$ can be derived via Eq.\,(\ref{EqScaSum}). For $\ell_\mathrm{an}$ and $\ell_\mathrm{is}$ the uncorrelated uncertainties are shown, while the second uncertainty given for $\ell_\mathrm{S}$ reflects the correlated effects, which mainly arise from the transmission factors $T$ of the greyfilters and polarizer, and from the relative photoefficiency $\varepsilon$ of the phototubes (Sect.\,\ref{SecDatAna}).\\
~\\
\textbf{PXE}\,\textsc{u}: For the unpurified sample, $\ell_\mathrm{an}$=34$\pm$4\,m is in good agreement with the expected value for $\ell_\mathrm{ray}$=32\,m (see below). $\ell_\mathrm{is}$=23$\pm$1\,m is the shortest isotropic scattering length of all measured samples. As the anisotropic scattering shows no sign of Mie contributions, it will mainly be due to absorption-reemission processes, $\ell_\mathrm{is}$$\approx$$\ell_\mathrm{are}$. This is consistent with the relatively low purity of 97\,$\%$ reported in the product specification sheet \cite{pxe-spec}. The main organic impurities are molecules closely related to regular PXE: PPXE and PMXE feature benzene rings and probably absorption bands at longer wavelengths. The resulting value $\ell_\mathrm{S}$=13.6$\pm$1.2\,m is therefore relatively short.\\
~\\
\textbf{PXE}\,\textsc{p}: A further measurement performed after purification of the PXE in an aluminum column shows an increase in the total scattering length to $\ell_\mathrm{S}$=22$\pm$3\,m. However, due to the low quality of the fit the data points had to be evaluated according to Eq.\,(\ref{EqMeaRes}); both scattering lengths $\ell_\mathrm{is}$=40$\pm$4\,m and $\ell_\mathrm{an}$=51$\pm$13\,m are greater than in the unpurified sample, but feature large uncertainties.\\
~\\
\textbf{LAB:} $\ell_\mathrm{an}$$\approx$40$\pm$5\,m is similar for all samples and slightly lower than the calculated value $\ell_\mathrm{ray}$=45\,m. $\ell_\mathrm{is}$$\approx$67$\pm6$\,m is quite large; this corresponds to the expected low level of contamination with organic impurities. These are mostly n-decanes which are not supposed to lower the transparency (compare the results for C12). Concerning $\ell_\mathrm{S}$$\approx$25$\pm2$\,m, no significant difference can be found between the three different brands.\\
~\\
\textbf{C12:} As expected, dodecane is the most transparent of the tested solvents. Both the samples from Sigma Aldrich and Alway Chem show only small isotropic contributions: $\ell_\mathrm{is}$$\approx$130\,m or even larger. Also for $\ell_\mathrm{an}$$\approx$45\,m, no large differences between the products from Sigma Aldrich and Alway Chem can be found. The total scattering length  $\ell_\mathrm{S}$$\approx$35$\pm$4\,m is the longest of the three solvents, reflecting the high transparency of C12.\\
~\\
\textbf{PC} is the solvent used in the currently running Borexino and KamLAND experiments (in the  second case under the addition of n-paraffins) \cite{bx08det,kam04}. Its optical properties have been thoroughly investigated in \cite{bx00sci}: With $\ell_\mathrm{S}$=7.8$\pm$0.8\,m the present work reproduces the former result. In accordance to the expectation, $\ell_\mathrm{an}$=19$\pm$3\,m$\approx$$\ell_\mathrm{ray}$ is rather short due to the high density of scattering centers. Like in the case of PXE\,\textsc{p}, mean values of the ratio $Q$ were used in the evaluation as the reduced $\chi^2$ of the fit is rather high.\\
~\\
\textbf{CX} can be used to cross-check the validity of the measured results. CX is supposed to show no significant absorption for $\lambda>150$\,nm \cite{sow72}. This is well reproduced in the measurements of this highly pure sample: The inverse of $\ell_\mathrm{is}$, $Q_\mathrm{is}$=(8$\pm$64)$\times$10$^{-5}$\,m$^{-1}$, is compatible with 0. The measured anisotropic scattering length $\ell_\mathrm{an}$=$\ell_\mathrm{S}$=45$\pm$5\,m is in excellent agreement with the predictions for $\lambda_\mathrm{ray}$=44\,m \cite{kay04,sow72,wah95}. CX can therefore be seen as a benchmark for the reliability of the other measurements; $\ell_\mathrm{an}$ is very sensitive to the correct treatment of the background measurement, especially the used Rayleigh scattering length of water (Sect.\,\ref{SecDatAna}).\\
~\\
The Rayleigh scattering lengths $\ell_\mathrm{ray}$ that are quoted for comparison have been derived  using the total Rayleigh cross section
\begin{eqnarray}\label{EqRayLam}
\sigma_\mathrm{ray}(\lambda,\lambda_0) = \frac{8\pi}{3}r_e^2 \frac{ {\lambda_0}^4} {\left(\lambda^2-{\lambda_0}^2 \right)^2}   \stackrel{\lambda\gg\lambda_0}{\longrightarrow} \frac{8\pi}{3}r_e^2\frac{{\lambda_0}^4}{\lambda^4}.
\end{eqnarray}

Here, $r_e$=2.9\,fm is the classic electron radius, while $\lambda_0$ is the resonance wavelength which corresponds to the spectral absorption maximum of the solvent molecule. For long wavelengths, $\ell_\mathrm{ray}(\lambda)$ can be approximated by the well-known $\lambda^4$ dependence \cite{hulst}.  

For PXE, LAB, and PC, the expected Rayleigh length can be estimated using Eqs.\,(\ref{EqScaLen}) and (\ref{EqRayLam}). The necessary absorption maxima and number densities have been adopted from \cite{pxe-spec,lab-spec,mar08phd,pc51}. For CX, former measurements at longer wavelengths can be extrapolated to 430\,nm \cite{kay04,sow72,wah95}.

In general, both Rayleigh and Mie scattering will contribute to the anisotropic scattering component. However, $\ell_\mathrm{ray}$$\approx$$\ell_\mathrm{an}$ seems to hold rather well for all samples, leaving only small room for Mie scattering. Moreover, for none of the samples an asymmetry between forward and backward scattering amplitudes could be found, which - if present - would also be a strong indication for Mie scattering (Sect.\,\ref{SecLigPro}). As neither asymmetry nor a surplus of anisotropic scattering is observed, the measurements point toward a negligible contribution of Mie scattering. Thus, $\ell_\mathrm{is}$ can be directly associated with the absorption-reemission length $\ell_\mathrm{are}$.

\begin{table}
\begin{center}
\begin{tabular}{lccccc}
\hline
Sample & $\ell_\mathrm{is}$ [m] & $\ell_\mathrm{an}$ [m] & $\ell_\mathrm{S}$ [m] & $\chi^2$/ndf & $\ell_\mathrm{ray}$\\
\hline
PXE\,\textsc{u}	&	22.8$\pm$1.0	&	33.6$\pm$4.0		&	13.6$\pm$0.7$\pm$1.0	&	1.39		& 32\\
PXE\,\textsc{p}  &	40.0$\pm$3.9	&	51$\pm$13		&	22.3$\pm$2.7$\pm$1.6	&	3.71		& 32\\   
C12\,\textsc{sa}	&	258$\pm$54	&	40.9$\pm$3.9		&	35.3$\pm$3.0$\pm$2.2	&	0.92		& \\
C12\,\textsc{ac}	&	132$\pm$16	&	48.5$\pm$5.6		&	35.4$\pm$3.1$\pm$2.3	&	0.77		& \\
LAB\,\textsc{p500}&	75.3$\pm$5.3	&	40.2$\pm$4.4		&	26.2$\pm$1.9$\pm$1.6	&	1.23		& 45\\
LAB\,\textsc{p550}&	60.5$\pm$3.7	&	40.5$\pm$5.2		&	24.3$\pm$1.9$\pm$1.5	&	1.29		& 45\\
LAB\,\textsc{550q}&	66.3$\pm$5.7	&	40.0$\pm$4.6		&	25.0$\pm$1.9$\pm$1.6	&	0.80		& 45\\
PC				&	13.0$\pm$0.9	&	19.3$\pm$3.3		&	7.8$\pm$0.6$\pm$0.6	&	1.52		& 21\\
CX				&	$>$10$^3$	&	45.0$\pm$4.5		&	44.9$\pm$4.5$\pm$2.9	&	0.74		& 44\\
\hline
\end{tabular}
\caption{Results from the measurement series performed at 430\,nm. The isotropic $\ell_\mathrm{is}$, anisotropic $\ell_\mathrm{an}$, and the total scattering length $\ell_\mathrm{S}$ are reported. For $\ell_\mathrm{is}$ and $\ell_\mathrm{an}$, only the uncorrelated uncertainties are quoted, for $\ell_\mathrm{S}$ also the correlated uncertainties. The reduced $\chi^2$-value of the analysis fit are shown as well. The expected Rayleigh scattering length $\ell_\mathrm{ray}$ is shown for comparison (Sect.\,\ref{SubsRes430}).}
\label{TabSerOne}
\end{center}
\end{table}  

\subsection{Wavelength Dependence}

\begin{table}
\begin{center}
\begin{tabular}{lccccc}
\hline
Sample 	& $\lambda$ [nm]	& $\ell_\mathrm{is}$ [m] & $\ell_\mathrm{an}$ [m] 	& $\ell_\mathrm{S}$ [m] 	& $\chi^2$/ndf\\
\hline
PXE\,\textsc{u}	& 415	&	14.0$\pm$0.8			&	27.8$\pm$5.2			&	9.3$\pm$0.6$\pm$0.7	&	0.52	\\
				& 430	&	18.6$\pm$1.4			&	32.1$\pm$7.1			&	11.8$\pm$1.0$\pm$0.9	&	1.86	\\
				& 442	&	21.1$\pm$0.4			&	41.7$\pm$9.5			&	14.0$\pm$1.2$\pm$1.1	&	0.92	\\
\hline
PXE\,\textsc{p}	& 415	&	37.5$\pm$2.9			&	65$\pm$17			&	23.8$\pm$2.5$\pm$1.8	&	0.72	\\
				& 430	&	40.0$\pm$3.9			&	51$\pm$13			&	22.3$\pm$2.7$\pm$1.6	&	3.71	\\
\hline
C12\,\textsc{sa} & 415	&	158$\pm$32			&	36.9$\pm$4.9			&	29.9$\pm$3.4$\pm$2.3	&	0.60	\\
				& 430	&	133$\pm$29			&	49$\pm$10			&	36.0$\pm$5.6$\pm$2.6	&	2.62	\\
				& 442	&	381$\pm$149			&	76$\pm$18			&	63$\pm$13$\pm$5		&	0.82	\\
\hline
PC 				& 415 	&	11.4$\pm$0.7			&	19.6$\pm$3.2			&	7.2$\pm$0.5$\pm$0.6	&	0.51	\\
				& 430	&	13.0$\pm$0.9			&	19.3$\pm$3.3			&	7.8$\pm$0.6$\pm$0.6	&	1.52	\\
				& 442	&	17.0$\pm$1.1			&	33.8$\pm$7.2			&	11.3$\pm$0.9$\pm$0.9	&	0.37	\\
\hline
CX 				& 415	&	$>$10$^3$			&	41.1$\pm$5.4			&	44.2$\pm$6.4$\pm$3.4	&	0.33	\\
				& 430	&	$>$10$^3$			&	40.8$\pm$6.0			&	45.1$\pm$7.9$\pm$3.3	&	2.48	\\
				& 442	&	512$\pm$389			&	53.0$\pm$9.8			&	48.0$\pm$8.5$\pm$3.8	&	1.16	\\
\hline
\end{tabular}
\caption{Results from the second measurement series performed at three different wavelengths. Anisotropic $\ell_\mathrm{an}$, isotropic $\ell_\mathrm{is}$ and the resulting total scattering length $\ell_\mathrm{S}$ are reported for original and purified PXE, C12, PC, and CX. The last column shows the reduced $\chi^2$-value of the analysis fit.}
\label{TabSerTwo}
\end{center}
\end{table} 

C12, PXE (both original \textsc{'u'} and purified \textsc{'p'}), and PC (the solvent currently used in the Borexino experiment) were tested along with CX. The latter two were included to provide a standard for comparison with former measurements. 

To monitor the wavelength dependence of the individual scattering components, the narrow bandpass in front of the sample was changed to perform measurements at 415, 430, and 442\,nm (Sect.\,\ref{SecExpSet}). The resulting scattering lengths are shown in Tab.\,\ref{TabSerTwo}. The measured scattering lengths of PC and CX are in good agreement with former measurements \cite{bx00sci,kay04,sow72,wah95}.

\begin{figure}
\centering
\includegraphics[width=0.48\textwidth]{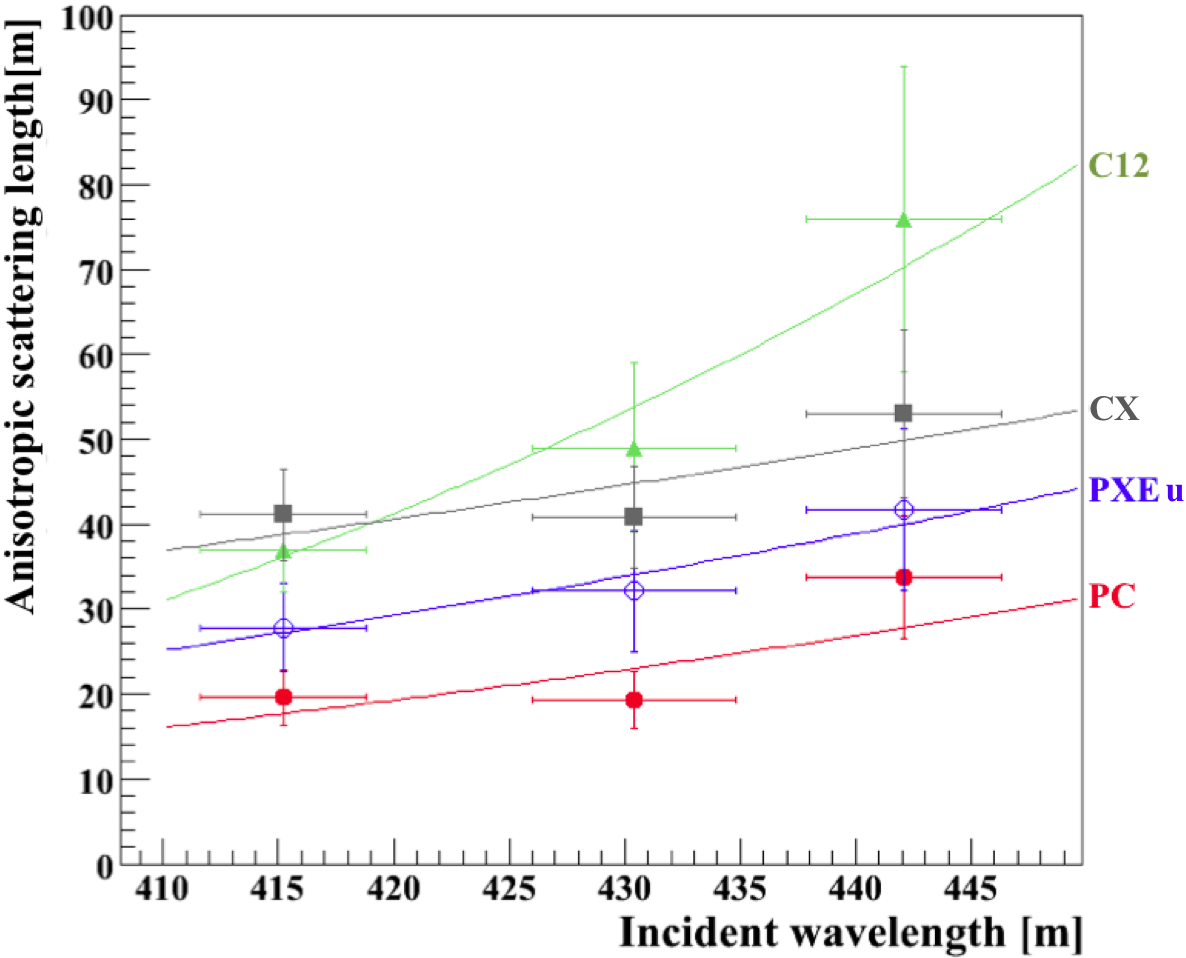}
\caption{Results for the anisotropic scattering length $\ell_\mathrm{an}$ for PXE\,\textsc{'u'}, C12, PC, and cyclohexane (CX) at three different wavelengths. For comparison, fits according to Eq.\,(\ref{EqRayLam}) are shown as solid curves of the same color. The data points are in good agreement with the $\lambda^4$-dependence expected for Rayleigh scattering.}
\label{FigRayLen}
\end{figure}

Fig.\,\ref{FigRayLen} shows the measured anisotropic scattering lengths $\ell_\mathrm{an}$ of PXE\,\textsc{u}, PC, C12, and CX for the three wavelength. The expected increase in scattering length with wavelength is observed for all samples. In addition, fits using the $\lambda$-dependence of Eq.\,(\ref{EqRayLam}) are shown. The best fit parameters as well as the $\chi^2$-values are listed in Tab.\,\ref{TabFitPar}. Especially for PC and PXE\,\textsc{'u'}, the absorption maximum $\lambda_0$ is in good agreement with expectations. However, the uncertainties for $\lambda_0$ are large.

In general, the quality of the wavelength-dependent measurements is lower than the results obtained for the fixed wavelength in Sect.\,\ref{SubsRes430}. This is correlated to a degradation of the glass surface of the beaker and probably also to the frequent changes of bandpasses and LED light intensity that decrease the stability of the LED emission. Nevertheless, reasonable values can be obtained when using the mean values of $Q_\perp$ and $Q_\parallel$ as described in Sect.\,\ref{SecDatAna}. These results are quoted in Tab.\,\ref{TabSerTwo} along with the reduced $\chi^2$ values to give an indication of the general quality of the data point. Note, however, that the results for 430\,nm featuring the worst $\chi^2$ values are in good agreement with the measurements of the same samples in the first series.

As PC and purified PXE were only characterized as part of the wavelength-dependent measurement, the results presented in Tab.\,\ref{TabSerOne} are identical to the values at 430\,nm in Tab.\,\ref{TabSerTwo}. Unfortunately, PXE\,\textsc{p} is affected the most by the afore-mentioned degradation of the glass surfaces. While the 415\,nm data point is reliable, the one at 430\,nm should be used with care, and the data of the 442\,nm measurement was not usable and therefore omitted from Tab.\,\ref{TabSerTwo}.

\begin{table}
\begin{center}
\begin{tabular}{|l|cc|ccc|}
\hline
 & \multicolumn{2}{c|}{Expected} & \multicolumn{2}{c}{Fit Result} & \\
Sample & $\lambda_0$\,[nm] & $\ell_\mathrm{ray}$\,[m] & $\lambda_\mathrm{0}$\,[nm] & $\ell_\mathrm{ray}$\,[m] & $\chi^2$/ndf \\
\hline
PXE\,\textsc{u}	& 265	& 32 & 252$\pm$192	& 34$\pm$4	& 0.11 \\
C12\,\textsc{sa}	& 		& 	 & 338$\pm$47		& 53$\pm$8	& 0.28 \\
PC				& 265	& 21 & 285$\pm$147	& 23$\pm$3	& 1.99 \\
CX				& $<$120	& 44 & 0$\pm$650		& 45$\pm$4	& 0.68 \\
\hline
\end{tabular}
\caption{Expected values \cite{pxe-spec,kay04,sow72,wah95,lab-spec,mar08phd,pc51} and best-fit parameters resulting from a fit of Eq.\,(\ref{EqRayLam}) to the measured anisotropic scattering length at three different wavelengths. The values of the absorption-band maximum $\lambda_0$ agree rather well with predictions, while uncertainties are quite large. $\ell_\mathrm{ray}$ corresponds to the Rayleigh scattering length at 430\,nm. $\chi^2$ indicates the quality of the fit (ndf=1).}
\label{TabFitPar}
\end{center}
\end{table}  

\section{Discussion and Conclusions}
\label{SecConclu}

The scattering lengths of organic solvents characterized in the present work vary between $\sim$8\,m for PC and 35\,m for C12 in the 430\,nm wavelength band. The expected values for the Rayleigh scattering lengths and the measured results are in good agreement. The $\lambda^4$-dependence of the Rayleigh scattering length is found. The proportion of absorption/re-emission processes can be clearly correlated to the chemical purity of the samples. Evidence of Mie scattering is not observed.

Scintillation and light transport in a large-volume liquid-scintillator detector are a complex interplay of several wavelength-dependent parameters of solvents, solutes, and photomultipliers. It is difficult to evaluate the importance of a single quantity, e.\,g., the scattering length outside the context of a solvent's overall performance, and almost impossible to provide accurate predictions for the performance of a final detector without a fair knowledge of all relevant parameters and the use of extensive Monte Carlo simulations of the setup.

Nevertheless, the measured scattering lengths allow a few general assertions concerning the aptitude of the investigated solvents for future large-scale detectors of masses of 10\,kt and more. While PC is an excellent choice for present 100\,t to 1\,kt scale detectors, the rather short scattering length will most likely lead to a rapid deterioration of its capabilities in vertex and time reconstruction with increasing detector dimensions. However, energy resolution will still be excellent, as light absorption processes are almost non-existent in PC \cite{bx00sci}.

The situation for PXE is comparable: Without purification by column chromatography, the scattering length is very low. In addition, a large impact of light absorption was visible in attenuation length measurements \cite{bx04pxe}, which would mean a significant reduction of the light output in a large detector. Purification or, alternatively, chemically cleaner production of PXE would offer a viable candidate for a detector of 20 to 30\,m diameter. Measurements show that attenuation length $L$=10\,m or slightly larger can be reached \cite{bx04pxe}. The value $\ell_\mathrm{S}$=22\,m obtained for the scattering length in this work indicates that scintillation signals from the bulk of the detector should not be smeared out too much in time before they reach the photomultipliers. Nevertheless, absorption would be of the same order as the scattering processes, reducing the effective light yield considerably in comparison to, e.\,g., Borexino \cite{bx07be7,bx08det,mar05}.

However, diluting either PC or PXE with C12 or other n-decanes might be a good option to increase the optical transparency of the liquid in order to minimize both scattering and absorption. Laboratory experiments in the frame of R\&D activities for DoubleChooz as well as the experience with KamLAND show that adding up to 80\,\% of C12 only moderately decreases the initial scintillation light yield (by 10-20\,\%) \cite{dc04,kam04}, while the gain in attenuation length might more than compensate these losses. Recently, values of $L$ of several tens of meters at 430\,nm have been reported from spectroscopic measurements \cite{buc09}. 

Based on the present results as well as R\&D measurements performed for the future SNO+ experiment, LAB seems to be the best solution for a large-volume detector. Without further purification, $L$=15-20\,m \cite{buc09,SNO08phd} and $\ell_\mathrm{S}$=25\,m reduce the light losses during propagation to the PMTs to a minimum. The smearing of the arrival time pattern due to scattering will have only a minor impact on the time resolution as long as the diameter of the scintillator tank does not exceed 30 to 40\,m. However, for certain applications, e.\,g.\,nucleon decay search, the relatively slow fluorescence decay of LAB (as determined in \cite{mar09}) in comparison to PC/PXE might imply a drawback as the efficiency for resolving fast coincidences is reduced \cite{mar05}.

The present knowledge on solvent parameters suggests LAB or mixtures of PC/PXE and C12 as the most viable candidates for liquid-scintillator detectors of 10\,kt mass or above. While additional experiments, e.\,g.\,a precision measurement of the wavelength-dependent attenuation length in a scintillator cell of several meters in length, are still missing, detector diameters of 30\,m or more seem feasible. The present results can serve as a valuable input for upcoming simulations of the scintillation light transport in large-scale detectors. 
~\\

\section*{Acknowledgements}

This work has been supported by the Maier-Leibnitz-Laboratorium (Garching), the Deutsche Forschungsgemeinschaft DFG (Transregio 27: Neutrinos and Beyond), and the Munich Cluster of Excellence (Origin and Structure of the Universe).

\bibliographystyle{h-physrev}

\end{document}